\documentclass[12pt]{article}

\begin{document}

\title{Equations of Motion in Kaluza-Klein Gravity Reexamined}
\author{J. Ponce de Leon\thanks{E-mail: jponce@upracd.upr.clu.edu or jpdel@astro.uwaterloo.ca}\\ Laboratory of Theoretical Physics, Department of Physics\\ 
University of Puerto Rico, P.O. Box 23343, San Juan, \\ PR 00931, USA}

\maketitle

\begin{abstract}
We discuss the equations of motion of test particles for a version of Kaluza-Klein theory where the cylinder condition is not imposed. The metric tensor of the five-dimensional manifold is allowed to depend on the fifth coordinate. This is the usual working scenario in brane-world, induced-matter theory and other Kaluza-Klein theories with large extra dimensions. We present a new version for the fully covariant splitting of the $5D$ equations. We show how to change the usual definition of various physical quantities in order to make physics in $4D$ invariant under transformations in $5D$. These include the redefinition of the electromagnetic tensor, force and Christoffel symbols. With our definitions, each of the force terms in the equation of motion is gauge invariant and orthogonal to the four-velocity of the particle. The ``hidden" parameter associated with the rate of motion along the extra dimension is identified with the electric charge, regardless of whether there is an electromagnetic field or not. 
In addition, for charged particles, the charge-to-mass ratio should vary. 
Therefore, the motion of a charged particle should differ from the motion of a  neutral particle, with the same initial mass and energy, even in the absence of electromagnetic field. 
These predictions have important implications and could in principle be experimentally detected. \end{abstract}

 
\newpage

\section{INTRODUCTION}

Kaluza's great achievement was the discovery that extending the number of dimensions from four to five allows the unification of gravity and electromagnetism. He showed that the five-dimensional Einstein equations, in vacuum, contain four-dimensional general relativity in the presence of an electromagnetic field, together with Maxwell's theory of electromagnetism. (There is also a Klein-Gordon equation for a massless scalar field that was suppressed at that time by adopting $g_{44}=constant$). The appearance of the ``extra" dimension in physical laws was avoided by imposing the ``cylinder condition", which essentially requires that all derivatives with respect to the fifth coordinate vanish.  
 
There are three versions of Kaluza theory \cite{Overduin}. The first one is known as compactified Kaluza-Klein theory. In this approach, Kaluza's cylinder condition is explained through a physical mechanism for compactification of the fifth dimension proposed by Klein. In the second version this condition is explained using projective geometry, in which the fifth dimension is absorbed into ordinary four-dimensional spacetime provided the (affine) tensors of general relativity are replaced with projective ones. 

In the third version the cylinder condition is not imposed and there are no assumptions about the topology of the fifth dimension. This is the usual scenario in induced-matter theory \cite{JPdeL 1}-\cite{JPdeL 3}, brane-world \cite{Arkani1}-\cite{RS2} and other non-compact Kaluza-Klein theories \cite{Overduin}, which assume that our four-dimensional spacetime is embedded in a world with more that four large dimensions. 

The study of the motion of particles provides in principle a way of testing whether there are extra dimensions to spacetime of the sort proposed by any of the abovementioned approaches. This requires the study of predictions of various theories and confrontation with experiment. In particular, with experiments involving the classical tests of relativity.
 
The equation of motion of a particle in $5D$ has been studied by a number of people. Much of the work was based on compactified versions of Kaluza-Klein theory, where there is no dependence of the metric on the extra (or internal) coordinate \cite{Leibowitz}-\cite{Gegenberg}. 
The corresponding equation for the third version, where the metric is allowed to depend on the extra coordinate, has also been derived \cite{Mashhoon}-\cite{Seahra2}. This equation is fully covariant in $4D$ and contains some terms that depend on the extra dimension. It has been discussed in a number of physical situations \cite{Wesson}. Despite successful applications, this equation presents two particular features that we regard as deficiencies. They are:
 
\medskip
 
(i) The force terms are not invariant under a group of transformations that we call gauge transformations. 

(ii) The associated ``fifth" force has a component parallel to particle's four-velocity. 

\medskip

These two features will be discussed in the next Section. Our aim in this work is to provide a new version for the $4D$ equation of motion, in which all force terms are gauge invariant and orthogonal to particle's four-velocity.

The plan for the paper is as follows. In the next section we discuss the features mentioned above. In Section 3, we present our splitting technique and notations. In Section 4, we do the splitting of the $5D$ geodesic. We obtain the ``generalized" electromagnetic tensor and ``projected" Christoffel symbols. We also show how to obtain the equations from a Lagrangian. In Section 5 we define and derive the appropriate fifth force. In Section 6, we discuss the initial value problem and the interpretation of the equations. In Section 7 we discuss some  experimental/observational implications of our formulation, i.e. we address the question of how can one distinguish the present formulation from general relativity from an experimental point of view. Finally, in Section 8, we summarize our results.

\section{Statement of The Problem}
The $5D$ line element is taken in the form
\begin{equation}
\label{5D metric in special coordinates}
d{\bf {\cal S}}^{2} = ds^{2} + \epsilon \Phi^{2}(dx^{4} + A_{\mu}
dx^{\mu})^{2},
\end{equation}
 where $ds^{2} = g_{\mu \nu} dx^{\mu} dx^{\nu}$ is the spacetime interval, while $\Phi$ and $A_{\mu}$ are the scalar and vector potentials. All these quantities, are functions of $x^{\mu}$ and the extra coordinate $x^{4}$. The factor $\epsilon$ is taken to be $+1$ or $-1$ depending on whether the extra dimension is timelike or spacelike, respectively. The $5D$ equations of motion are obtained by minimizing interval (1). From them, the equations for a test particle moving in ordinary $4D$ are taken as \cite{Wesson}
 \begin{equation}
\label{poor slicing, with respect to dS}
\frac{d^2 x^{\mu}}{d{\cal S}^2}+ {\Gamma}^{\mu}_{\alpha \beta}\frac{dx^{\alpha}}{d{\cal S}}\frac{dx^{\beta}}{d{\cal S}}= n F^{\mu}_{\;\;\nu}\frac{dx^{\nu}}{d{\cal S}}+ 
\epsilon n^2 \frac{\Phi^{;\mu}}{\Phi^{3}}    - A^{\mu}\frac{dn}{d{\cal S}}- g^{\mu\lambda}\frac{dx^4}{d{\cal S}}\left(n
 \frac{\partial{A_{\lambda}}}{\partial{x^4}}+\frac{\partial{g_{\lambda\nu}}}{\partial{x^4}}\frac{dx^{\nu}}{d{\cal S}}\right),
\end{equation}
and 
\begin{eqnarray}
\label{poor slicing}
\frac{d^2 x^{\mu}}{ds^2}&+& {\Gamma}^{\mu}_{\alpha \beta}\frac{dx^{\alpha}}{ds}\frac{dx^{\beta}}{ds}= \frac{n}{\left(1-\epsilon{n^2}/{\Phi^2}\right)^{1/2}}\left[F^{\mu}_{\;\;\nu}\frac{dx^{\nu}}{ds} - \frac{A^{\mu}}{n}\frac{dn}{ds}- g^{\mu\lambda}\frac{\partial{A_{\lambda}}}{\partial{x^4}}\frac{dx^4}{ds}   \right]+\nonumber \\ & &\frac{{\epsilon} {n}^2}{\left(1 -\epsilon{n}^2/\Phi^2\right)\Phi^3}\left[\Phi^{;\mu} + \left(\frac{\Phi}{n}\frac{dn}{ds}- \frac{d\Phi}{ds}\right)\frac{dx^{\mu}}{ds}\right]-g^{\mu\lambda}\frac{\partial{g_{\lambda\nu}}}{\partial{x^4}}\frac{dx^{\nu}}{ds}\frac{dx^4}{ds}   
\end{eqnarray}
where ${\Gamma}^{\mu}_{\alpha\beta}$ represents the Christoffel symbol constructed from $g_{\mu\nu}$, $F_{\mu\nu}$ is the usual antisymmetric tensor
\begin{equation}
\label{Usual EM tensor}
F_{\mu\nu} = A_{\nu,\mu}-A_{\mu,\nu},
\end{equation}
and 
\begin{equation}
\label{equation for n}
n = \epsilon {\Phi}^2\left(\frac{dx^4}{d{\cal S}} + A_{\mu}\frac{dx^{\mu}}{d{\cal S}}\right).
\end{equation}
\subsection{The Problem}
It is clear that physics in $4D$ should be invariant under the set of transformations
\begin{eqnarray}
\label{Allowed transformations}
x^{\mu} &=& \bar{x}^{\mu}(x^\lambda),\nonumber \\ x^{4}&=& \bar{x}^{4}+ f(\bar{x}^{0},\bar{x}^{1},\bar{x}^{2},\bar{x}^{3}),
\end{eqnarray}
that keep invariant the given $(4+1)$ splitting. Indeed, in $5D$ they just reflect the freedom in the choice of origin for $x^{4}$, while in $4D$ they correspond to the usual gauge freedom of the potentials 
\begin{equation}
\label{gauge transformations}
\bar{A}_{\mu} = A_{\mu} + \frac{\partial{f}}{\partial{\bar{x}^{\mu}}}= A_{\mu}+ f,_{\mu}.
\end{equation}
The $5D$ interval (\ref{5D metric in special coordinates}), the spacetime metric $g_{\mu\nu}$ and the scalar field $\Phi$ remain invariant under these transformations, as one expects. However, their derivatives do change as 
\begin{eqnarray}
\label{transformation of spacetime metric}
\bar{g}_{\mu\nu,\lambda}&=& g_{\mu\nu,\lambda} + g_{\mu\nu,4}f_{,\lambda}\;, \nonumber  \\ 
\bar{\Gamma}^{\lambda}_{\alpha\beta}&=& {\Gamma}^{\lambda}_{\alpha\beta} + \frac{1}{2}g^{\lambda\rho}(g_{\rho\alpha,4}f_{,\beta} + g_{\rho\beta,4}f_{,\alpha} - g_{\alpha\beta,4}f_{,\rho})\;, \nonumber \\
\bar{\Phi}_{,\mu}&=& \Phi_{,\mu}+ \Phi_{,4}f_{,\mu}.
\end{eqnarray}
Also  
\begin{eqnarray}
\label{EM Tensor}
\bar{A}_{\mu,\nu} &=& A_{\mu,\nu} + A_{\mu,4}f_{,\nu} + f_{,\mu,\nu},\nonumber \\
\bar{F}_{\mu\nu}&=& F_{\mu\nu} + \left(A_{\nu,4}f_{\mu}- A_{\mu,4}f_{,\nu}\right).
\end{eqnarray}
These equations show that none of the forces (gravitational, scalar or Lorenz force)  in (\ref{poor slicing, with respect to dS}) or (\ref{poor slicing}), neither their combination, remains invariant under gauge transformations. Indeed, direct substitution of (\ref{gauge transformations})-(\ref{EM Tensor}) into the right-hand side of (\ref{poor slicing, with respect to dS}) or (\ref{poor slicing}), yields a combination of additional terms (that are of products of $f_{,\mu}$ with $g_{\mu\nu,4}$, $A_{\mu,4}$ or $\Phi_{,4}$) which do not cancel out, in general. In fact, the only way to make them vanish is to require total independence of the extra variable.

Thus, in the case where the metric functions depend on $x^{4}$, the Lorenz and gravitational ``force"  per unit mass as given by (\ref{poor slicing, with respect to dS}) and (\ref{poor slicing}) are gauge dependent. We regard this property as a deficiency of equations (\ref{poor slicing, with respect to dS}) and (\ref{poor slicing}). In this work we will construct a new version of the  $4D$ equation of motion in which each force term, separately, is gauge invariant and orthogonal to the four-velocity. 
\section{The Splitting Technique and Notation}

One of the great advantages of general relativity is the freedom in the choice of coordinate system. However, in many cases, this makes the coordinates to be merely marking parameters, without much physical content \cite{Landau}. This is a potential source for misinterpretation.
The coordinates $x^{\mu}$, in the $(4+1)$ separation given by (\ref{5D metric in special coordinates}), are spacetime coordinates, in the sense that $dx^{\mu}$ is an infinitesimal  displacement in $4D$. However, the change of a physical quantity along ``$x^{\mu}$ direction" is not given by $(\partial/\partial{x^{\mu}})$, in general. For this to be so, there should be no dependence on the extra variable at all. This is the source of the problems in (\ref{poor slicing, with respect to dS}) and (\ref{poor slicing}), as we learn from our equations (\ref{transformation of spacetime metric}) and (\ref{EM Tensor}).

In order to overcome these problems, we will start by considering a general five-dimensional manifold, with an arbitrary set of marking parameters, and will construct the physical quantities in $4D$, step by step. Then we will define the ``local" frame of reference that we will use through this work.

 \subsection{$4D$ Spacetime From $5D$}

Let us consider a general five-dimensional manifold with coordinates $\xi^{A}$ $(A = 0,1,2,3,4)$ and metric tensor $\gamma_{AB}(\xi^{C})$. The $5D$ interval is then given by
\begin{equation}
\label{general 5D metric}
d{\cal S}^2 = \gamma_{AB}d{\xi}^A d{\xi}^B.
\end{equation}
We should assume that this $5D$ manifold allows us to construct, appropriately (see bellow), a four-dimensional hypersurface that can be identified with our $4D$ spacetime. In this hypersurface we introduce a set of four parameters $x^{\mu}$ $(\mu= 0,1,2,3)$, which are functions of $\xi^A$,
\begin{equation}
\label{4D coordinates}
x^{\mu}= x^{\mu}\left({\xi^0},{\xi^1},{\xi^2},{\xi^3},{\xi^4}\right).
\end{equation}
The derivatives of these functions with respect to $\xi^A$
\begin{equation}
\label{Basis in 4D}
\hat{e}^{(\mu)}_{A}= \frac{\partial{x^{\mu}}}{\partial{\xi^A}},
\end{equation}
behave as covariant vectors\footnote{The index in parenthesis numbers the vector, while the other one indicates its coordinate in $5D$.} with respect to changes $\xi^A = {\xi^A}(\bar{\xi}^B)$ in $5D$, and as  contravariant vectors with respect to transformations $x^{\mu}= x^{\mu}(\bar{x}^{\nu})$ in $4D$. At each point these vectors are tangent to the hypersurface. Therefore, in the region where they are linearly independent, they constitute a basis for the $4D$ hypersurface under consideration. We will interpret this, appropriately defined $4D$ manifold, as the physical spacetime and $x^{\mu}$ as the coordinates in it.

We can now introduce the vector $\psi^{A}$, orthogonal to spacetime. This is completely determined by 
\begin{eqnarray}
\label{definition of psi}
\hat{e}^{(\mu)}_{A}\psi^{A}&=&0,\nonumber \\
\gamma_{AB}\psi^A\psi^B &=& \epsilon,
\end{eqnarray}
where $\epsilon$ is retained by the same reasons as in (\ref{5D metric in special coordinates}). In order to define projected quantities, we will also need the set of vectors $\hat{e}_{(\mu)}^{A}$, defined  as
\begin{eqnarray}
\label{associate basis}
\hat{e}_{A}^{(\mu)}\hat{e}^{A}_{(\nu)}&=& \delta^{\mu}_{\nu},\nonumber \\     
\psi_{A}\hat{e}^{A}_{(\mu)}&=& 0.
\end{eqnarray} 
It is not difficult to show that $\hat{e}^{A}_{(\mu)}$ behave as contravariant vectors\footnote{In general, vectors $\hat{e}^{A}_{\mu}$ are not partial derivatives of any function of spacetime coordinates $x^{\alpha}$.} with respect to changes $\xi^A = {\xi^A}(\bar{\xi}^B)$ in $5D$, and as  covariant vectors with respect to transformations $x^{\mu}= x^{\mu}(\bar{x}^{\nu})$ in $4D$. This follows from (\ref{associate basis}) and the transformation properties of $\hat{e}^{(\mu)}_{B}$.

Now, any five-dimensional vector, say $P_{A}$, can be split into two parts; a $4D$ part $P_{(\mu)} = \hat{e}^{A}_{(\mu)}P_{A}$ and a part parallel to $\psi^{A}$, namely $P_{(4)}= P_{A}\psi^{A}$. The $4D$ projection\footnote{$P_{\mu}$ is the $\mu$ component of the $5D$ vector $P_{A}$, while $P_{(\mu)}$ is the projection of $P_{A}$ in the direction of basis vector $\hat{e}^{A}_{(\mu)}$}, $P_{(\mu)}$, behaves like a covariant vector under general transformations in spacetime $x^{\mu}= x^{\mu}(\bar{x}^{\alpha})$ and it is invariant under transformations $\xi^A = \xi^{A}(\bar{\xi}^B)$ in $5D$.

The same can be done with partial derivatives. For example, the derivative $V_{\mu,A}$ contains two parts: a $4D$ part $V_{\mu|(\lambda)}= V_{\mu,A}\hat{e}^{A}_{(\lambda)}$, and a part orthogonal to it, which is $V_{\mu|(4)}= V_{\mu,A}\psi^{A}$. Thus\footnote{Projected derivatives are denoted by a ``$\mid$", followed by the direction of projection.},
\begin{equation}
\label{projected derivatives}
V_{\mu,A}= V_{\mu|(\rho)}\hat{e}^{(\rho)}_{A}+ \epsilon V_{\mu|(4)}\psi_{A}.
\end{equation} 
In particular, any infinitesimal displacement in $5D$ can be written as
\begin{equation}
\label{5D displacement}
d\xi^A = \hat{e}^{A}_{(\mu)}dx^{(\mu)}+ \epsilon \psi^A dx^{(4)},
\end{equation}
 where $dx^{(\mu)}= \hat{e}_{B}^{(\mu)}d\xi^B$ and $dx^{(4)} = \psi_{B}d\xi^B$ represent the displacements along the corresponding basis vectors. Substituting (\ref{5D displacement}) into (\ref{general 5D metric}) we obtain 
\[
d{\cal S}^2= \gamma_{AB}\hat{e}^{A}_{(\mu)}\hat{e}^{B}_{(\nu)}dx^{(\mu)}dx^{(\nu)}+ \epsilon \left(dx^{(4)}\right)^2.\]
 Consequently, the metric of the spacetime is given by
\begin{equation}
\label{4D metric}
g_{\mu\nu}= \hat{e}_{(\mu)}^{A}\hat{e}_{(\nu)}^{B}\gamma_{AB}.
\end{equation}
We also notice the consistency relation
\begin{equation}
\label{projector}
\hat{e}_{(\mu)}^{A}\hat{e}^{(\mu)}_{B}= \delta_{B}^{A}- \epsilon \psi^{A}\psi_{B},
\end{equation}
which follows from the above separation in $4D$ and scalar quantities.
\subsection{Local Frame}
Thus, in the neighborhood of each point the observer is armed with five independent vectors, $\hat{e}^{(\mu)}_{A}$ and $\psi^{B}$, that constitute its frame of reference. In order to simplify further calculations we introduce a more symmetrical notation. To this end we set 
\begin{eqnarray}
\label{extra basis vector}
\hat{e}^{(4)}_{A}&=& \psi_{A},\nonumber  \\
\hat{e}^{A}_{(4)}&=& \epsilon \psi^A.
\end{eqnarray}
Now, Eqs. (\ref{definition of psi}), (\ref{associate basis}) and (\ref{projector}) become
\begin{eqnarray}
\label{general relation between basis vectors}
\hat{e}_{A}^{(B)}\hat{e}^{A}_{(C)}&=& \delta_{C}^{B},\nonumber  \\
\hat{e}^{(N)}_{C}\hat{e}_{(N)}^{D}&=& \delta_{C}^{D}.
\end{eqnarray}
Finally, similar to Eq. (\ref{4D metric}), we define our ``local" $5D$ metric $\hat{g}_{(A)(B)}$ as
\begin{equation}
\label{Local 5D metric}
\hat{g}_{(A)(B)}= \hat{e}_{(A)}^{M}\hat{e}_{(B)M}.
\end{equation}
which breaks up the five-dimensional manifold, viz.,
\begin{equation}
\hat{g}_{(A)(B)}= \bordermatrix{  &  &      \cr
                                  &g_{\mu\nu} &0    \cr
                                  &0 &\epsilon    \cr}.
\end{equation}
The advantage of the local frame is that it provides a $(4+1)$ separation which is fully invariant under arbitrary changes of coordinates in $5D$, not only under the special class of transformations defined by (\ref{Allowed transformations}).
The $5D$ interval becomes
\begin{equation}
\label{5D interval in local metric}
d{\cal S}^2 = \hat{g}_{(A)(B)}dx^{(A)}dx^{(B)}.
\end{equation}
In addition, from (\ref{general relation between basis vectors}), it follows that 
\begin{equation}
\label{coordinate metric from local metric}
\gamma_{AB}= \hat{e}_{A}^{(C)}\hat{e}_{B}^{(D)}\hat{g}_{(C)(D)}.
\end{equation}
Finally, we mention that basis indexes are lowered and raised with $ \hat{g}_{(A)(B)}$, while  $5D$ coordinate indexes are lowered and raised with $\gamma_{AB}$.

\section{Splitting The $5D$ Geodesic}

The plan of this Section is as follows. First, we do the $(4 + 1)$ splitting of the $5D$ geodesic in an arbitrary local basis. Second, we apply the general results to a particular frame, that we call coordinate frame. Finally, we show how to simplify the splitting procedure using the Lagrangian formalism. 
\subsection{Arbitrary Frame}
 
By minimizing the interval (\ref{general 5D metric}) we obtain the $5D$ geodesic equation in covariant form
\begin{equation}
\label{5geod}
\frac{1}{2}(\gamma_{AB,C} + \gamma_{CB,A}- \gamma_{AC,B})\frac{d \xi^{A}}{d {\cal S}}\frac{d \xi^{C}}{d{\cal S}}+\gamma_{AB}\frac{d^{2}\xi^{A}}{d{{\cal S}}^{2}} = 0.
\end{equation}
Our task is to express this equation in terms of projected quantities. To obtain fully covariant equations, we will use our local metric (\ref{Local 5D metric}). First notice 
\begin{eqnarray}
\label{coordinate velocity and acceleration}
\frac{d{\xi}^A}{d{\cal S}}&=& \hat{e}^{A}_{(B)} U^{(B)},\nonumber  \\
\frac{d^2{\xi}^A}{d{\cal S}^2}&=& \hat{e}^{A}_{(B)|(P)}U^{(B)}U^{(P)} + \hat{e}^{A}_{(B)}\frac{dU^{(B)}}{d{\cal S}},
\end{eqnarray}
where 
\begin{eqnarray}
\label{5D velocity}
U^{(A)}&=& \frac{dx^{(A)}}{d{\cal S}},\nonumber \\
\hat{g}_{(A)(B)}U^{(A)}U^{(B)} &=& 1,
\end{eqnarray}
is the projected $5D$ velocity.
Now, we use (\ref{coordinate metric from local metric}) to express ``coordinate" metric $\gamma_{AB}$ in terms of local metric, and the orthogonality conditions (\ref{general relation between basis vectors}) to simplify some derivatives of the basis vectors. With this, and substituting (\ref{coordinate velocity and acceleration}) into (\ref{5geod}), after some algebra, we find
\begin{equation}
\label{projected 5D geod}
\frac{1}{2}\left( \hat{g}_{(Q)(N)|(P)}+\hat{g}_{(P)(N)|(Q)}-\hat{g}_{(P)(Q)|(N)}\right)U^{(P)}U^{(Q)}+ \hat{g}_{(N)(P)}\frac{dU^{(P)}}{d{\cal S}} = U_{(A)}{\cal F}^{(A)}_{(N)(P)}U^{(P)},
\end{equation}
where ${\cal F}^{(A)}_{(N)(P)}$ is defined as
\begin{equation}
\label{EM tensor generalized}
{\cal F}_{(N)(P)}^{(A)}= \hat{e}^{(A)}_{Q}(\hat{e}_{(N)|(P)}^{Q} - \hat{e}_{(P)|(N)}^{Q})
\end{equation}
The antisymmetric nature of this quantity remind us of the electromagnetic tensor. For this interpretation, however, the contravariant index requires a closer examination. Let us study the $A=\lambda$ components of this tensor. Using orthogonality conditions (\ref{general relation between basis vectors}) in (\ref{EM tensor generalized}) we obtain
\begin{equation}
\label{space part of Generalized EM tensor}
{\cal F}^{(\lambda)}_{(A)(B)} = \hat{e}^{P}_{(B)} \hat{e}^{(\lambda)}_{P|(A)}- \hat{e}^{P}_{(A)}\hat{e}^{(\lambda)}_{P|(B)}.
\end{equation}
We now need to remember that $\hat{e}^{(\mu)}_{A}= (\partial{x^\mu}/{\partial{\xi^A}})$. Using this, and since $({\partial}^2/{\partial{\xi}^P}{\partial{\xi^Q}})= ({\partial}^2/{\partial{\xi}^Q}{\partial{\xi^P}})$, it follows that
\[
{\cal F}^{(\lambda)}_{(A)(B)} = 0. \]
Therefore only ${\cal F}^{(4)}_{(A)(B)}$ survives. This antisymmetric tensor provides $10$ degrees of freedom. We will see that six of them are associated with the electromagnetic field, while the other four with the so called fifth force. In what follows the index ``(4)" in ${\cal F}^{(4)}_{(A)(B)}$ will be suppressed.   
  
The spacetime part of (\ref{projected 5D geod}) is obtained by setting\footnote{This is the projection on the spacetime basis vectors $\hat{e}^{(\mu)}_{A}$. This projection is invariant under general transformations in $5D$.}
 $N=\mu$
\begin{equation}
\label{4D from 5D}
\frac{dU^{(\mu)}}{d{\cal S}} + \hat{\Gamma}^{\mu}_{\alpha\beta}U^{(\alpha)}U^{(\beta)}= U_{(4)}{\cal F}_{(\lambda)(P)}U^{(P)}g^{\lambda\mu}- g^{\mu\lambda}g_{\lambda\nu|(4)}U^{(\nu)}U^{(4)},
\end{equation}
where we have raised the free index, and ${\hat{\Gamma}}^{\mu}_{\alpha\beta}$ is the Christoffel symbol constructed with the projected derivatives,
\begin{equation}
\label{projected Christoffel symbol}
\hat{\Gamma}^{\mu}_{\alpha\beta}= \frac{1}{2}g^{\mu\rho}\left(g_{\rho\alpha|(\beta)}+ g_{\rho\beta|(\alpha)} - g_{\alpha\beta|(\rho)}\right).
\end{equation}
This, ``projected" Christoffel symbol is invariant under general coordinate transformations in $5D$.  

We will see in Section 4.3 that the momentum, per unit mass, projected on our local frame is given by $p_{(N)}= \hat{g}_{(N)(P)}U^{(P)}$. Then from (\ref{projected 5D geod}), it follows that
\begin{equation}
\label{General Projected Momentum}
\frac{dp_{(N)}}{d{\cal S}}= \frac{1}{2} {g}_{\mu\nu|(N)}U^{(\mu)}U^{(\nu)} + U_{(4)}{\cal F}_{(N)(P)}
U^{(P)}.
\end{equation}
We will use this equation in our discussion of the fifth force, in the next Section. 

The above set of equations constitutes the basis for our further discussion. 
They are, by construction, covariant under transformations of coordinates in $4D$, and invariant under general transformations in $5D$.

The conclusion from the above discussion is as follows. In the local frame we calculate the projected Christoffel symbols $\hat{\Gamma}^{\mu}_{\alpha\beta}$. They constitute the appropriate affine connection to be used  when calculating covariant derivatives in $4D$ (otherwise, there will be no gauge invariance as in (\ref{poor slicing, with respect to dS}) and (\ref{poor slicing})). Thus, the usual gravitational ``force" in $4D$ will be invariant under transformations in $5D$. Then we calculate the antisymmetric tensor ${\cal F}_{(A)(B)}$, whose ten independent components (we will see) are related to the Lorentz and ``scalar" force. These forces are proportional to $U^{(4)}$.

\subsection{Coordinate Frame}

We now apply our general equations to the particular frame used in (\ref{5D metric in special coordinates}). With this aim, let us then consider the special case where
\begin{equation}
\label{special coordinates}
x^{\mu}= {\xi}^{\mu}.
\end{equation}
The spacetime basis vectors are
\begin{eqnarray}
\label{spacetime basis vectors}
\hat{e}^{(0)}_{A}&=& (1, 0, 0, 0, 0),\nonumber \\
\hat{e}^{(1)}_{A}&=& (0, 1, 0, 0, 0),\nonumber \\
\hat{e}^{(2)}_{A}&=& (0, 0, 1, 0, 0),\nonumber \\
\hat{e}^{(3)}_{A}&=& (0, 0, 0, 1, 0).
\end{eqnarray}
From (\ref{definition of psi}) we find
\begin{equation}
\label{e4}
\hat{e}^{A}_{(4)}= \epsilon{\psi^A}= (0, 0, 0, 0, \frac{\epsilon}{\Phi}),
\end{equation}
where we have set $\gamma_{44}= \epsilon \Phi^{2}$. The associated basis vectors are given by (\ref{general relation between basis vectors}). Denoting $\gamma_{\mu 4}= \epsilon \Phi^2 A_{\mu}$, we obtain
\begin{eqnarray}
\label{associated basis vectors}
\hat{e}^{A}_{(0)}&=& (1, 0, 0, 0, -A_{0}),\nonumber \\ 
\hat{e}^{A}_{(1)}&=& (0, 1, 0, 0, -A_{1}),\nonumber \\ 
\hat{e}^{A}_{(2)}&=& (0, 0, 1, 0, -A_{2}),\nonumber \\ 
\hat{e}^{A}_{(3)}&=& (0, 0, 0, 1, -A_{3}),\nonumber \\ 
\hat{e}^{(4)}_{A}&=& \epsilon \Phi(A_{0}, A_{1}, A_{2}, A_{3}, 1).
\end{eqnarray}
The $5D$ line element and the $4D$ metric become 
\begin{eqnarray}
\label{special metric} 
d{\cal S}^2 &=& g_{\mu\nu}dx^{\mu}dx^{\nu}+ \epsilon \Phi^2\left(d{\xi}^4 + A_{\mu}dx^{\mu}\right)^2,\nonumber \\
g_{\mu\nu}&=& \gamma_{\mu\nu}-\epsilon\Phi^2A_{\mu}A_{\nu}.
\end{eqnarray}
The interval shows the same separation as in (\ref{5D metric in special coordinates}), as one expected\footnote{Under transformation (\ref{Allowed transformations}); $\bar{\gamma}_{\mu\nu} = \gamma_{\mu\nu} + \epsilon\Phi^{2}(A_{\mu}f_{,\nu} + A_{\nu}f_{,\mu} + f_{,\mu}f_{,\nu})$ and $\bar{A}_{\mu} = (A_{\mu} + f_{,\mu})$, but the metric remains invariant $\bar{g}_{\mu\nu} = g_{\mu\nu}$.}. We will keep the use of $\xi^{4}$, in order to avoid any confusion with the ``physical" displacement along the extra dimension\footnote{In this frame, spacetime displacements are $dx^{\mu}$, while the ones along the extra dimension are $dx^{(4)}= \hat{e}^{(4)}_{A}d{\xi^A}= \epsilon \Phi(d{\xi}^4 + A_{\mu}dx^{\mu})$.}.
Also, 
\begin{eqnarray}
\label{Gen of EM tensor in special coordinates}
{\cal F}_{(\mu)(\rho)}&=& \epsilon \Phi\left(A_{\rho|(\mu)}- A_{\mu|(\rho)}\right),\nonumber \\
{\cal F}_{(\mu)(4)}&=& \frac{\Phi_{|(\mu)}}{\Phi}- \epsilon \Phi A_{\mu|(4)}.
\end{eqnarray} 
Finally, we substitute these expressions into the (\ref{4D from 5D}) and obtain the desired equation, viz., 
\begin{equation}
\label{spacetime projection of the $5D$ geodesic}
\frac{dU^{(\sigma)}}{d{\cal S}} + \hat{\Gamma}^{\sigma}_{\alpha\beta}U^{(\alpha)}U^{(\beta)}= n\hat{F}^{\sigma}_{\;\; \rho}U^{(\rho)}+ \epsilon n^2 \frac{\Phi^{|(\sigma)}}{\Phi^3}- g^{\sigma\mu}U^{(4)}\left(nA_{\mu|(4)}+ g_{\mu\rho|(4)}U^{(\rho)}\right),
\end{equation}
where $n= {\Phi}U^{(4)}$ is the same scalar as in (\ref{poor slicing, with respect to dS}), 
\begin{eqnarray}
\label{projected derivatives in the special frame}
\Phi^{|(\sigma)}&=& g^{\sigma\lambda}\Phi_{|(\lambda)} = g^{\sigma\lambda}\left(\Phi_{,\lambda}- \Phi_{,4}A_{\lambda}\right),\nonumber  \\
g_{\mu\rho|(4)}&=& {\epsilon} \frac{g_{\mu\rho,4}}{\Phi},\nonumber \\
A_{\mu|(4)}&=&{\epsilon}\frac{A_{\mu,4}}{\Phi},
\end{eqnarray}
and  
\begin{eqnarray}
\label{hatF}
\hat{F}_{\mu\rho}&=& \left(A_{\rho|(\mu)}-A_{\mu|(\rho)}\right)\nonumber \\
&=& \left(A_{\rho,\mu}-A_{\mu,\rho}\right) + \left(A_{\rho}A_{\mu,4}-A_{\mu}A_{\rho,4}\right).
\end{eqnarray}
We will see in Section 6 that this quantity, instead of (\ref{Usual EM tensor}), plays the role of ``generalized" electromagnetic tensor in the present $5D$ theory. The above equation is invariant under ``gauge" transformations (\ref{Allowed transformations}). The use of $\hat{\Gamma}^{\mu}_{\alpha\beta}$ guarantees the gauge invariance of the gravitational force. Also, the above defined $\hat{F}_{\mu\rho}$ is gauge invariant. Consequently, the Lorenz force is invariant too. The same is true for the ``scalar"  force associated with $\Phi_{|\mu}$. 

We conclude, from the above discussion, that equation (\ref{spacetime projection of the $5D$ geodesic}) should replace that in (\ref{poor slicing, with respect to dS}). It is not the equation of motion yet, because the later involves differentials with respect to $ds$ instead of $d{\cal S}$, and we still have to do the splitting of $(dU^{(\sigma)}/d{\cal S})$ in a ``$4+1$" parts. We will discuss this in Section 5 too.
 
\subsection{Lagrangian Method}
 Equation (\ref{projected 5D geod}) gives the components of $5D$ geodesics on an arbitrary set of basis vectors $\hat{e}^{(A)}_{B}$. Its derivation from (\ref{5geod}) is straightforward, but involves some tedious calculations. On the other hand, the geodesic equation, in its general form (\ref{5geod}), is obtained in a very simple, direct, way from the Lagrangian density 
\begin{equation}
\label{Lagrangian density}
L= \frac{1}{2}\gamma_{AB}{\dot{\xi}}^A{\dot{\xi}}^B,
\end{equation}
where $\dot{\xi}^A = d{\xi}^A/d{\cal S}$, as usual. Therefore the question arises whether it is possible to obtain the components of this equation, on a given local frame $\hat{e}^{(A)}_{B}$, right away from a Lagrangian. The answer to this question is positive. To do this one should use our local metric. Indeed, substituting (\ref{coordinate metric from local metric}) into (\ref{Lagrangian density}) we get
\begin{equation}
\label{Local Lagrangian density }
L= \frac{1}{2}{\hat{g}}_{(M)(N)}\hat{e}^{(M)}_{A}\hat{e}^{(N)}_{B}{\dot{\xi}}^A{\dot{\xi}}^B.
\end{equation}
Now, taking the derivatives $({\partial{L}/{\partial{\xi}^A}})$ and $({\partial{L}/{\partial{\dot{\xi}}^A}})$ and using the Lagrangian equation
\begin{equation}
\label{Lagrangian equation}
\frac{d}{d{\cal S}}\left(\frac{\partial{L}}{\partial{{\dot{\xi}}}^A}\right) - \frac{\partial{L}}{\partial{{{\xi}}^A}}= 0,
\end{equation}
we readily obtain (\ref{projected 5D geod}). In addition to this, we get the appropriate definitions for the generalized momentum per unit mass $P_{A}$, viz.,
\begin{equation}
\label{generalized momentum}
P_{A}= \frac{\partial{L}}{\partial{{\dot{\xi}}}^A} = \hat{g}_{(M)(N)}U^{(M)}\hat{e}^{(N)}_{A}.
\end{equation}
From which we get its components on our local frame. They are, 
\begin{equation}
\label{projected momentum}
p_{(A)}= \hat{g}_{(A)(B)}U^{(B)}.
\end{equation}
The equation governing $p_{(A)}$ was already obtained in (\ref{General Projected Momentum}), while for the generalized momentum it is
\begin{equation}
\label{evolution of generalized momentum}
\frac{dP_{C}}{d{\cal S}}= \frac{1}{2}g_{\mu\nu,C}U^{(\mu)}U^{(\nu)} + \left(g_{\mu\nu}U^{(\mu)}\hat{e}^{(\nu)}_{B,C}+ U_{(4)}\hat{e}^{(4)}_{B,C}\right){\dot{\xi}}^B,
\end{equation} 
which can be obtained either from $P_{C}= p_{(A)}\hat{e}^{(A)}_{C}$, or from the ``local" Lagrangian (\ref{Local Lagrangian density }). In the case where the metric is independent of $\xi^{C}$, the corresponding component of the generalized momentum is a constant of motion\footnote{The 0-component, as well as the 4-component can be written as, $\dot{P}_{0}= (\gamma_{AB,0}\dot{\xi}^{A}\dot{\xi}^B/2)$ and $\dot{P}_{4}= (\gamma_{AB,4}\dot{\xi}^{A}\dot{\xi}^B/2)$, respectively.}. 

In the coordinate frame the generalized momentum, per unit mass, is given by
\begin{eqnarray}
\label{gen. momentum in coordinate frame}
P_{\lambda}&=& g_{\mu\lambda}U^{(\mu)} + n A_{\lambda},\nonumber  \\
P_{4}&=& n.
\end{eqnarray}
Its components on the local frame are
\begin{eqnarray}
\label{momentum in local frame}
p_{(\lambda)}&=&g_{\mu\lambda}U^{(\mu)},\nonumber  \\
p_{(4)}&=& \epsilon \frac{n}{\Phi}.
\end{eqnarray}

\section{The Equation of Motion in $4D$} 

We now proceed to obtain the equations of motion in $4D$. Our plan of action is as follows. First, we find the absolute derivatives of the four-velocity. Second, we show that the straightforward extension, of the definition of force used in $4D$ general relativity, to evaluate the fifth force leads to some problems. Then, we proceed to split the absolutes derivatives and introduce a more appropriate definition for the fifth force in $4D$.  

The four-velocity is defined as usual
\begin{eqnarray}
\label{four-velocity}
u^{(\mu)}&=& \frac{dx^{(\mu)}}{ds},\nonumber \\
g_{\mu\nu} u^{(\mu)}u^{(\nu)} &=& 1.
\end{eqnarray}
From (\ref{5D interval in local metric}) we obtain the relation between $d{\cal S}$ and $ds$, viz.,
\begin{equation}
\label{interval, new version}
d{\cal S}^2 = \hat{g}_{(A)(B)}dx^{(A)}dx^{(B)} =
  g_{\mu\nu}dx^{(\mu)}dx^{(\nu)}+ \epsilon(dx^{(4)})^2 
 = ds^2 + \epsilon(dx^{(4)})^2.
\end{equation}
Consequently,
\begin{equation}
\label{relation between ds and dS}
d{\cal S} = ds \sqrt{1 + \epsilon \left(\frac{dx^{(4)}}{ds}\right)^2}.
\end{equation}
In order to have a more ``symmetrical" notation, in what follows we set
\begin{equation}
\label{u4}
u^{(4)}= \frac{dx^{(4)}}{ds}.
\end{equation}
To avoid misunderstanding, we stress the fact that $u^{(4)}$ is not a part of the four-velocity vector $u^{(\mu)}$.
Now, using (\ref{5D velocity}), we find  
\begin{equation}
\label{relation between four and five velocity}
 U^{(A)}= \left( \frac{u^{(\mu)}}{\sqrt{1 + \epsilon (u^{(4)})^2}}\;\;\;, \;\;\; \frac{u^{(4)}}{\sqrt{1 + \epsilon (u^{(4)})^2}}\right).
\end{equation}

\subsection{Absolute Derivative of Four-Velocity} 
On the local frame, the spacetime components of the momentum (per unit mass) are given by (\ref{momentum in local frame}). Thus,
\begin{eqnarray}
\label{d/dS of spacetime components of the momentum}
\frac{dp_{(\mu)}}{d{\cal S}}&=& \frac{d}{d{\cal S}}(g_{\mu\nu}U^{(\nu)})=\frac{d}{d{\cal S}}\left(g_{\mu\nu}\frac{u^{(\nu)}}{\sqrt{1 + \epsilon (u^{(4)})^2}}\right)\nonumber \\
&= & \frac{1}{{\sqrt{1 + \epsilon (u^{(4)})^2}}}\frac{d}{ds}\left(\frac{u_{(\mu)}}{{\sqrt{1 + \epsilon (u^{(4)})^2}}}\right)\nonumber  \\
&=& \frac{1}{\left[{{1 + \epsilon (u^{(4)})^2}}\right]}\frac{du_{(\mu)}}{ds}-\frac{{\epsilon}u_{(\mu)}{u^{(4)}}}{\left[{{1 + \epsilon (u^{(4)})^2}}\right]^2} \frac{du^{(4)}}{ds}.
\end{eqnarray}
 Setting $N = 4$ in (\ref{projected 5D geod}) we obtain 
 \begin{equation}
\label{N = 4 projected geod}
\epsilon\frac{dU^{(4)}}{d{\cal S}}= \frac{1}{2}g_{\mu\nu|(4)}U^{(\mu)}U^{(\nu)}+ \epsilon{\cal F}_{(4)(P)}U^{(4)}U^{(P)},
\end{equation}
from which we get
\begin{equation}
\label{equation for u4}
\frac{\epsilon}{\left[{{1 + \epsilon (u^{(4)})^2}}\right]}\frac{du^{(4)}}{ds}= \frac{1}{2}g_{\mu\nu|(4)}u^{(\mu)}u^{(\nu)}+ \epsilon{\cal F}_{(4)(\rho)}u^{(4)}u^{(\rho)}.
\end{equation}
We now substitute this expression into (\ref{d/dS of spacetime components of the momentum}) and obtain
\begin{equation}
\label{dP/dS from N = 4}
\frac{dp_{(\mu)}}{d{\cal S}}= \frac{1}{\left[{{1 + \epsilon (u^{(4)})^2}}\right]}\left[\frac{du_{(\mu)}}{ds} - u_{(\mu)}u^{(4)}\left(\frac{1}{2}g_{\lambda\rho|(4)}u^{(\lambda)}u^{(\rho)}+ \epsilon{\cal F}_{(4)(\rho)}u^{(4)}u^{(\rho)}\right)\right].
\end{equation}
On the other hand, from (\ref{General Projected Momentum}) we have
\begin{equation}
\label{General Projected Momentum with N = mu}
\frac{dp_{(\mu)}}{d{\cal S}}= \frac{1}{\left[{{1 + \epsilon (u^{(4)})^2}}\right]}\left(\frac{1}{2}g_{\lambda\rho|(\mu)}u^{(\lambda)}u^{(\rho)} + \epsilon u^{(4)}u^{(P)}{\cal F}_{(\mu)(P)}\right).
\end{equation}
Equating the last two expressions we obtain
\begin{eqnarray}
\frac{du_{(\mu)}}{ds}&=& \frac{1}{2}u^{(\alpha)}u^{(\beta)}\left(g_{\alpha\beta|(\mu)} + u_{(\mu)}g_{\alpha\beta|(4)}u^{(4)}  \right)
+ (\epsilon u^{(4)}){\cal F}_{(\mu)(\rho)}u^{(\rho)}\nonumber  \\  &+& \epsilon \left(u^{(4)}\right)^{2} \left({\cal F}_{(\mu)(4)} + u_{(\mu)}{\cal F}_{(4)(\rho)}u^{(\rho)}  \right).
\end{eqnarray}
Now we notice that
\begin{equation}
\label{absolut differential}
\frac{Du_{(\mu)}}{ds}= \frac{du_{(\mu)}}{ds} - {\hat{\Gamma}}^{\tau}_{\mu\nu}u_{(\tau)}u^{(\nu)}=   \frac{du_{(\mu)}}{ds} -\frac{1}{2}g_{\alpha\beta|(\mu)}u^{(\alpha)}u^{(\beta)}.
\end{equation}
Consequently,
\begin{eqnarray}
\label{absolut differential of u covariant}
\frac{Du_{(\mu)}}{ds}= (\epsilon u^{(4)}){\cal F}_{(\mu)(\rho)}u^{(\rho)} &+& \epsilon (u^{(4)})^{2} \left({\cal F}_{(\mu)(4)} + u_{(\mu)}{\cal F}_{(4)(\rho)}u^{(\rho)}  \right)\nonumber  \\ &+& \frac{1}{2}u_{(\mu)}g_{\lambda\rho|(4)}u^{(\lambda)}u^{(\rho)}u^{(4)}.
\end{eqnarray}
In a similar way, from (\ref{4D from 5D}) and (\ref{equation for u4}) we obtain
\begin{eqnarray}
\label{absolut differential of u contravariant}
\frac{Du^{(\sigma)}}{ds}= (\epsilon u^{(4)}){\cal F}^{(\sigma)}_{\;\;\; (\rho)}u^{(\rho)} + \epsilon (u^{(4)})^2 \left({\cal F}^{(\sigma)}_{\;\;\; (4)} + u^{(\sigma)}{\cal F}_{(4)(\rho)}u^{(\rho)}\right)\nonumber \\ 
+ \frac{u^{(\sigma)}}{2}g_{\lambda\rho|(4)}u^{(\lambda)}u^{(\rho)}u^{(4)} - g^{\sigma\lambda}g_{\lambda\rho|(4)}u^{(\rho)}u^{(4)}.
\end{eqnarray}
\subsection{Definition of Force in the Literature}

As an extension of the concept of force in $4D$ general relativity \cite{Landau}, the extra (or ``fifth") force per unit mass acting on a particle is defined as \cite{Mashhoon}-\cite{Wesson},
\begin{equation}
\label{force in the literature}
f_{(lit)}^{\mu}= \frac{Du^{(\mu)}}{ds}.
\end{equation} 
Because this is a fully covariant $4D$ equation one would expect
\begin{equation}
f_{(lit)\sigma}= g_{\sigma\mu}f^{\mu}_{(lit)}= \frac{Du_{(\sigma)}}{ds}.
\end{equation}
However, as we can easily see from (\ref{absolut differential of u covariant}) and (\ref{absolut differential of u contravariant}), this is not so. Instead we have
\begin{equation}
\label{relation between covariant and contravariant components of force in literature}
f_{(lit)\mu}= g_{\mu\sigma}f_{(lit)}^{\sigma} + g_{\mu\rho|(4)}u^{\rho}u^{(4)}.
\end{equation}
When the metric is independent of the extra variable, the last term vanishes and we have the correct relation between the covariant and contravariant components of the force. However, this is not so, for the general case under consideration here. Therefore, the adoption of definition (\ref{force in the literature}) would lead to a theory where $(Du_{(\mu)}/ds)$ and $(Du^{(\mu)}/ds)$ would be the covariant and contravariant components of different vectors\footnote{There would be an ambiguity between the covariant and contravariant  components of $Du^{(\mu)}/ds$ and $Du_{(\mu)}/ds$.}. This is equivalent to taking away one of the most important properties of the metric tensor, which is to lower and raise indexes. Apart of this, the force $f^{\mu}_{lit}$ defined in (\ref{force in the literature}) has the peculiar property  of not being orthogonal to the four-velocity. All this of course means that the force defined by (\ref{force in the literature}) is not a four-vector. This conclusion was recently confirmed, using another formalism, by Seahra \cite{Seahra3}.   

\subsection{Splitting Absolute Derivatives}
Our viewpoint is that we do not need to change the properties of the $4D$ metric tensor, what we  need is a better definition for the force. In order to do that, let us examine the absolute differential in more detail. Consider any $4D$ geometrical object, for the sake of the argument, let say a vector $V_{\alpha}$. Then,
\begin{eqnarray}
\label{splitting absolute differential}
DV_{\alpha}&=& dV_{\alpha} - \hat{\Gamma}^{\lambda}_{\alpha\rho}V_{\lambda}dx^{(\rho)}\nonumber \\ 
&=& \left(V_{\alpha|(\rho)} - \hat{\Gamma}^{\lambda}_{\alpha\rho}V_{\lambda}\right)dx^{(\rho)} +  V_{\alpha|(4)}dx^{(4)}.
\end{eqnarray}
The absolute differential separates into two parts, viz.,
\begin{equation}
\label{separation of D}
DV_{\alpha}= D^{(4)}V_{\alpha} + V_{\alpha|(4)}dx^{(4)},
\end{equation}
where $D^{(4)}$ represents the absolute differential in $4D$, namely
\begin{equation}
\label{D4 of vector V}
D^{(4)}V_{\alpha}= \left(V_{\alpha|(\rho)} - \hat{\Gamma}^{\lambda}_{\alpha\rho}V_{\lambda}\right)dx^{(\rho)}.
\end{equation}
 This separation is invariant under transformations in $5D$, provided all derivatives are projected appropriately and $\hat{\Gamma}^{\lambda}_{\alpha\rho}$ is that defined in (\ref{projected Christoffel symbol}).
Obviously, for any object we have
\begin{equation}
\label{def of D in 4D}
D^{(4)}(\cdot\cdot\cdot)= D(\cdot\cdot\cdot) - (\cdot\cdot\cdot)_{|(4)}dx^{(4)}.
\end{equation} 
In particular, for the metric tensor
\begin{equation}
\label{D4 of metric tensor}
D^{(4)}g_{\mu\nu}= \left[g_{\mu\nu|(\rho)} - \left(\hat{\Gamma}^{\lambda}_{\mu\rho}g_{\lambda\nu}+ \hat{\Gamma}^{\lambda}_{\nu\rho}g_{\lambda\mu}\right) \right]dx^{(\rho)} = 0,
\end{equation}
as it should be. 

\subsection{New Definition For The Fifth-Force}

In this paper we propose to define the fifth force (per unit mass) as follows
\begin{equation}
\label{correct definition of force}
f^{\mu}= \frac{D^{(4)}u^{(\mu)}}{ds}, \;\;\;\;\;\  f_{\mu}= \frac{D^{(4)}u_{(\mu)}}{ds},
\end{equation}
which, we believe, is in the original spirit of $4D$. With this definition the metric tensor preserves its property of lowering and raising indexes. Indeed, because of (\ref{D4 of metric tensor}) we have $f_{\sigma} = g_{\sigma\mu}f^{\mu}$, as desired. 
\\
Let us now find the contravariant components, $f^{\mu}$. Since $D^{(4)}u^{(\mu)} = Du^{\mu} - u^{\mu}_{|(4)}dx^{(4)}$, we need to evaluate $u^{(\mu)}_{|(4)}$. 
\begin{eqnarray}
\label{calculation of partial derivative of u}
du^{(\mu)} = d{\left(\frac{dx^{(\mu)}}{ds}\right)}&=& \frac{d(dx^{(\mu)})}{ds}-\frac{dx^{(\mu)}}{(ds)^2}d(ds)\nonumber \\
&=&\frac{d(dx^{(\mu)})}{ds}-\frac{dx^{(\mu)}}{(ds)^2}d\left(\sqrt{g_{\alpha\beta}dx^{(\alpha)}dx^{(\beta)}}\right).
\end{eqnarray}
Taking derivatives and rearranging terms
\begin{equation}
\label{calculation of partial derivative of u, part 2}
du^{(\mu)}= \frac{d(dx^{(\mu)})}{ds} - \left(g_{\alpha\beta} u^{(\alpha)}\frac{du^{(\beta)}}{ds}\right)dx^{(\mu)} - \frac{1}{2}u^{(\mu)}g_{\alpha\beta,A}u^{(\alpha)}u^{(\beta)}d{\xi}^A. 
\end{equation}
From this, and using that $d\xi^A = \hat{e}^{A}_{(P)}dx^{(P)}$, we get\footnote{Another way of obtaining this result is using the comoving frame where $u^{(\mu)} = \delta^{\mu}_{0}/\sqrt{g_{00}}$.}
\begin{equation}
\label{calculation of partial derivative of u, part 3}
u^{(\mu)}_{|(4)}= - \frac{1}{2}u^{(\mu)}g_{\alpha\beta|(4)}u^{(\alpha)}u^{(\beta)}.
\end{equation}
For the covariant components $f_{\mu}$ we need $u_{(\mu)|(4)}$. This can be obtained from above and $u_{(\mu)} = g_{\mu\nu}u^{(\nu)}$, as
\begin{equation}
\label{partial derivative of u covariant}
u_{(\mu)|(4)} = g_{\mu\lambda|(4)}u^{(\lambda)} - \frac{1}{2}u_{(\mu)}g_{\alpha\beta|(4)}u^{(\alpha)}u^{(\beta)}.
\end{equation}
We now have everything we need to write the $4D$ equation of motion in appropriate form;
\begin{eqnarray}
\label{general expression for my definition of force, contravariant}
\frac{D^{(4)}u^{(\sigma)}}{ds} = f^{\sigma}= \epsilon u^{(4)}{\cal F}^{(\sigma)}_{\;\;\; (\rho)}u^{(\rho)} + \epsilon (u^{(4)})^2 \left[{\cal F}^{(\sigma)}_{\;\;\; (4)} + u^{(\sigma)}{\cal F}_{(4)(\rho)}u^{(\rho)}\right]\nonumber \\ 
+ \left[u^{(\sigma)}u^{(\lambda)} - g^{\sigma\lambda}\right]g_{\lambda\rho|(4)}u^{(\rho)}u^{(4)}.
\end{eqnarray}
Also,
\begin{eqnarray}
\label{general expression for my definition of force, covariant}
\frac{D^{(4)}u_{(\mu)}}{ds} = f_{\mu} = \epsilon u^{(4)}{\cal F}_{(\mu)(\rho)}u^{(\rho)} &+& \epsilon (u^{(4)})^{2} \left[{\cal F}_{(\mu)(4)} + u_{(\mu)}{\cal F}_{(4)(\rho)}u^{(\rho)}  \right]\nonumber  \\ &+& \left[u_{(\mu)}u^{(\rho)} - {\delta}^{\rho}_{\mu}\right]g_{\rho\lambda|(4)}u^{(\lambda)}u^{(4)}.
\end{eqnarray}
To these equations, we should add the one for the evolution of $u^{(4)}$, which is given by (\ref{equation for u4}). Also we notice that if the metric were independent of some of the coordinates, say $\xi^{A}$, then the conjugate component of the generalized momentum (\ref{generalized momentum}) would be constant of motion, 
\begin{equation}
\label{conjugate component of the generalized momentum there is a constant of motion}
P_{A}= \frac{1}{\sqrt{1 + \epsilon (u^{(4)})^2}}(g_{\mu\nu}u^{(\mu)}\hat{e}^{(\nu)}_{A} +\epsilon u^{(4)}\hat{e}^{(4)}_{A}). 
\end{equation}
 The above equations are totally general. Namely; (i) They are expressed in an arbitrary frame of basis vectors $\hat{e}^{(A)}_{B}$; (ii) They are invariant under general transformations in $5D$, not only the restricted set mentioned in (\ref{Allowed transformations}); (iii) They behave like $4D$ vectors under coordinate transformations $x^{\mu}= x^{\mu}(\bar{x}^{\nu})$; (iv) The metric tensor retains its property of raising and lowering indexes, and (v) The force is orthogonal to the four-velocity, i.e., $f_{\mu}u^{(\mu)}= f^{\sigma}u_{(\sigma)}= 0$. 

\subsection{Equations of Motion in Coordinate Frame}

Let us now specialize our choice of basis vectors. As in Section (4.2) we consider the frame defined by the vectors (\ref{spacetime basis vectors}). In this frame, the non-zero components of ${\cal F}_{(A)(B)}$  are given by (\ref{Gen of EM tensor in special coordinates}). Direct substitution in (\ref{general expression for my definition of force, covariant}) yields\footnote{Here we omit the brackets for the components of the four-velocity because, in this frame, the coordinate displacements coincide with those along the basis vectors. Also $u^{(4)}= (dx^{(4)}/ds) = \epsilon \Phi[(d{\xi}^{4}/ds) + A_{\mu}u^{\mu}$].}   
\begin{eqnarray}
\label{equation of motion in coordinate frame}
\frac{D^{(4)}u_{\mu}}{ds}& &= (\Phi u^{(4)})\hat{F}_{\mu\rho}u^{\rho} + \Phi (u^{(4)})^2\left[A_{\rho|(4)} u^{\rho}u_{\mu} - A_{\mu|(4)} \right]  +\nonumber  \\ 
& &  \frac{\epsilon (u^{(4)})^2}{\Phi}\left[\Phi_{|(\mu)} - u_{\mu}\Phi_{|(\rho)}u^{\rho} \right] + \left[u_{(\mu)}u^{(\rho)} - {\delta}^{\rho}_{\mu}\right]g_{\rho\lambda|(4)}u^{\lambda}u^{(4)},
\end{eqnarray}
where the projected derivatives\footnote{In this frame, the rule is as follows: $V_{|(4)}= \epsilon (V_{,4}/\Phi)$ and $V_{|(\mu)}= V_{,\mu} - V_{,4}A_{\mu}$.} and $\hat{F}_{\mu\nu}$ are given by equations (\ref{projected derivatives in the special frame}) and (\ref{hatF}), respectively. The left-hand side of this equation is the spacetime component of the absolute derivative calculated with the projected Christoffel symbols, and it is invariant with respect to gauge transformations. Therefore, it is perfectly identical to that in Einstein's theory. The force terms on the right-hand side are deviations from four-dimensional geodesic motion. Equation (\ref{equation of motion in coordinate frame}) is the correct $4D$ equation and should replace that in (\ref{poor slicing}).
\\
The equation for $u^{(4)}$ can be obtained from (\ref{equation for u4}), which now becomes
\begin{equation}
\label{equation for u4 in the coordinate frame}
\frac{\epsilon}{\left[{{1 + \epsilon (u^{(4)})^2}}\right]}\frac{du^{(4)}}{ds}= \frac{1}{2}g_{\mu\nu|(4)}u^{(\mu)}u^{(\nu)}+ \left[ \Phi A_{\mu|(4)}- \epsilon\frac{\Phi_{|(\mu)}}{\Phi} \right]u^{\mu}u^{(4)}.
\end{equation}
These equations are invariant under the set of gauge transformations (\ref{Allowed transformations}), which leave invariant the spacetime basis vectors $\hat{e}^{(\mu)}_{B}$. They constitute a system of five differential equations with five unknowns\footnote{We recall that $u^{(4)}$ is not a component of the four-velocity vector.}, namely, $u^{0}$, $u^{1}$, $u^{2}$, $u^{3}$ and $ u^{(4)}$. For a general  $5D$ metric, with full dependence on the extra coordinate, and $A_{\mu} \neq 0$, the solution and analysis of  
(\ref{equation of motion in coordinate frame}) and (\ref{equation for u4 in the coordinate frame}) would probably require the use of numerical calculations. Certain simplification would be attained if the metric were independent of some coordinate. In this case the corresponding component of the generalized momentum (\ref{gen. momentum in coordinate frame}) would be a constant of motion, viz.,
\begin{eqnarray}
\label{conservation of gen. momentum in coordinate frame} 
P_{\lambda} &=& \frac{1}{{\sqrt{1 + \epsilon (u^{(4)})^2}}}(g_{\lambda\rho}u^{\rho} + \Phi u^{(4)}A_{\lambda}),\nonumber \\
P_{4} &=& \frac{\Phi u^{(4)}}{{\sqrt{1 + \epsilon (u^{(4)})^2}}}.
\end{eqnarray}

\section{Interpretation of $u^{(4)}$}

In the case of no dependence on the extra coordinate our equation (\ref{equation of motion in coordinate frame}) correctly reproduce the same results obtained previously in compactified Kaluza-Klein theory \cite{Leibowitz}-\cite{Gegenberg}. Indeed, the terms inside the second bracket, as well as the last term, all vanish. In addition, $\hat{F}_{\mu\nu}$ reduces to the electromagnetic tensor $F_{\mu\nu}$  defined as usual in (\ref{Usual EM tensor}). In this case, the multiplicative term in front of the corresponding $\hat{F}_{\mu\nu}$ is identified with the charge-to-mass ratio, in such a way that the first term on the right-hand side of (\ref{equation of motion in coordinate frame}) is interpreted as the Lorenz force.   
\subsection{Usual Interpretation. Case $A_{\mu} \neq 0$}

We will extend this interpretation to our theory. Specifically, 
\[
\hat{F}_{\mu\rho} = \left(A_{\rho|(\mu)}-A_{\mu|(\rho)}\right)
 = \left(A_{\rho,\mu}-A_{\mu,\rho}\right) + \left(A_{\rho}A_{\mu,4}-A_{\mu}A_{\rho,4}\right)
\]
will be interpreted as the electromagnetic tensor in the Kaluza-Klein theory under consideration. Accordingly, will interpret the first term on the right-hand side of (\ref{equation of motion in coordinate frame}) as the ``generalized"  Lorentz force. Consequently, we can write 
\begin{equation}
\label{charge to mass ratio}
\frac{q}{m} = (\Phi u^{(4)}),
\end{equation}
for the charge-to-mass ratio of the test particle. Thus, in the presence of an electromagnetic field, we relate the electric charge to its rate of motion along the extra dimension. This is the usual interpretation.

\subsection{Further Interpretation. Case $A_{\mu} = 0$}
 We thus arrived at the question: How should we interpret $u^{(4)}$ in the absence of electromagnetic field?. 

 There is no consensus answer to this. In fact many authors just leave this quantity as a free parameter without interpretation \cite{Overduin},\cite{Wesson}. But let us imagine the following scenario: a charged particle moving in an electromagnetic field that dies off with time. According to the above interpretation, while the field is not zero we  would relate its electric charge to $u^{(4)}$ as in (\ref{charge to mass ratio}). Then, the question arises, should we abandon this interpretation as soon as the field dies off? Apparently, not. Because electric charge is an intrinsic property of the particle; it does not depend on how we switch on and off the electric field. Once the particle ``chooses" its local frame, 
the quantity $u^{(4)}= \hat{e}^{(4)}_{A}(d{\xi}^A/ds)$ is invariant under any transformation $\xi^A = \xi^A(\bar{\xi}^B)$ in $5D$. On the other hand, we can use this freedom to make $\gamma_{4\mu}= 0$ if we desire to switch off the electromagnetic field without changing $u^{(4)}$.

The proposal we consider here is that the electric charge of a particle is always related to its ``velocity" $u^{(4)}$, via (\ref{charge to mass ratio}), regardless of whether it is moving is an electromagnetic field or not. Besides the above-mentioned general ideas, we have some physical and mathematical reasons to consider such interpretation. 
\subsubsection{Initial Value Problem}
Let us consider a particle moving in a region without electromagnetic field, and assume $u^{(4)}$ {\bf is not} proportional to the charge. Then, like we mentioned earlier, equations (\ref{equation of motion in coordinate frame}) and (\ref{equation for u4 in the coordinate frame}) constitute a set of five differential equations of second order to calculate five unknowns\footnote{In absence of electromagnetic field there is no distinction between $x^{4}$ and $\xi^4$.}; namely, $x^{0}$, $x^{1}$, $x^{2}$, $x^{3}$ and $ x^{4}$. The complete specification of the solution requires the initial values of eight quantities (not ten because there are two constraints; $g_{\mu\nu}u^{\mu}u^{\nu} =1$ and $\gamma_{AB}U^{A}U^{B} = 1$). These can be taken as follows: the initial time $t_{0}$ , six quantities corresponding to the initial position ${\bf r}_{0} = (x^{1}_{0},x^{2}_{0},x^{3}_{0})$ and initial spatial velocity $\dot{\bf r}_{0} = (\dot{x}^{1}_{0},\dot {x}^{2}_{0},\dot{x}^{3}_{0})$, and the initial value $u^{4}_{0}$. The trajectory of the particle would be given by\footnote{Gauge transformations (\ref{Allowed transformations}) reflect the freedom in the choice of origin for $\xi^{4}$.} 
\begin{equation}
\label{sacrifice of predicting power}
{\bf r} = {\bf r}(t - t_{0}, {\bf r}_{0}, \dot{\bf r}_{0}, u^{4}_{0}).
\end{equation}
If this were the case, we would not be able to give a complete specification of the motion of a particle without knowing its initial (hidden) velocity along the extra dimension. Therefore, different particles\footnote{These are ``classical" (not quantum particles) because they have well defined position and velocity at the same time.} having identical initial position and velocity would move along different trajectories if they have different initial values for $u^{4}$. This situation is clearly illustrated by a test body in radial free fall near a soliton, where the velocity in the fifth dimension affects its rate of fall in a very significant way \cite{radial motion}. If this were indeed the case, we would have a classical theory with (almost) no predicting power. Therefore, either we provide an interpretation for $u^{4}$, or we sacrifice the predicting power of the theory. 
\subsection{Our Interpretation and Final Equations}
A possible way to save the predicting power of the theory (at this level) is to consider that $u^{(4)}$ is always related to the electric charge of the particle, regardless of whether there is an electromagnetic field or not. In this case, the $4D$ trajectory will be given by  
\begin{equation}
\label{restoring the predicting power}
{\bf r} = {\bf r}(t - t_{0}, {\bf r}_{0}, \dot{\bf r}_{0}, (q/m)_{0}).
\end{equation}
Thus, knowing the position, velocity and charge-to-mass ratio, at any given time, we are able to give a complete specification of the motion of a particle. This sounds more satisfactory from a point of view of classical physics. For this interpretation, the equation for the charge-to-mass ratio can be obtained from (\ref{equation for u4 in the coordinate frame}) as    
\begin{eqnarray}
\label{final expression for the charge to mass ratio}
\frac{d}{ds}\left(\frac{q}{m}\right)= \frac{1}{2}g_{\mu\nu,4}u^{\mu}u^{\nu} &+& \left(\frac{q}{m}\right)A_{\mu,4}u^{\mu} + \left(\frac{q}{m}\right)^2\left[\frac{\Phi_{,4}}{\Phi} + \frac{1}{2}g_{\mu\nu,4}u^{\mu}u^{\nu}\right]\frac{\epsilon}{\Phi^2}\nonumber \\
 &+& \left(\frac{q}{m}\right)^3\left[ A_{\mu,4} - \frac{\Phi_{|(\mu)}}{\Phi}\right]\frac{\epsilon u^{\mu}}{\Phi^2}.
\end{eqnarray}
The corresponding equation of motion becomes
\begin{eqnarray}
\label{equation of motion in terms of the ratio q/m}
\frac{D^{(4)}u_{\mu}}{ds} = \left(\frac{q}{m}\right)\left[\hat{F}_{\mu\rho}u^{\rho} + \left(u_{(\mu)}u^{(\rho)} - {\delta}^{\rho}_{\mu}\right)\Phi^{-1}g_{\rho\lambda|(4)}u^{\lambda}\right]\;\;\;\;\;\;\;\;\;\;\;\;\;\;\;\ \nonumber \\ 
+ \left(\frac{q}{m}\right)^2\left[\left(A_{\rho|(4)} u^{\rho}u_{\mu} - A_{\mu|(4)} \right)\Phi^{-1}  + 
 \epsilon \left(\Phi_{|(\mu)} - u_{\mu}\Phi_{|(\rho)}u^{\rho} \right)\Phi^{-3}\right].
\end{eqnarray}
The important feature of this system is that it contains no reference to quantities in $5D$. Elsewhere we will discuss these equations, for some particular metrics, in more detail.

A point of interest should be mentioned here. If we set $\epsilon = 0$ we erase the five-dimensional part of the metric; $d{\cal S} = ds$. Then putting all derivatives with respect to the extra coordinate equal to zero, we obtain $(q/m) = Constant$ from (\ref{final expression for the charge to mass ratio}), while from (\ref{equation of motion in terms of the ratio q/m}) we get the $4D$ geodesic with the Lorenz force. The electromagnetic field does not vanish in this limit.  

\subsection{Neutral Particles}

The effects of the extra dimensions can be most readily appreciated in the case of charged particles, because of the force term on the right-hand side of (\ref{equation of motion in terms of the ratio q/m}). 

In the case of neutral particles, setting $q = 0$ all terms on the right-hand side of (\ref{equation of motion in terms of the ratio q/m}) vanish. Therefore, the motion of neutral test particles is governed by the usual equation in $4D$ general relativity, viz.,
\begin{equation}
\label{usual equation of motion in 4D}
\frac{D^{(4)}u^{\mu}}{ds} = 0.
\end{equation}
In addition, from equation (\ref{final expression for the charge to mass ratio}) we get
\begin{equation}
\label{constraint equation}
g_{\mu\nu, 4}u^{\mu}u^{\nu} = 0.
\end{equation}
This does not imply $g_{\mu\nu,4}= 0$, in general. Instead, this is a bilinear combination between the components of the four-velocity, which should be taken as a constraint equation. This constraint has to be solved simultaneously with the $4D$ geodesic equation.  

However, it is not clear whether it is possible to solve the
geodesic equation subject to $g_{\alpha\beta,4} u^\alpha u^\beta =
0$ for an arbitrary (local) frame. On the other hand, given a five-dimensional metric (\ref{general 5D metric}) we have the freedom to choose the set of $4D$ coordinates (\ref{4D coordinates}) as we wish. Except for mathematical simplicity, there are no criteria for this choice. In particular, there are no {\em physical} reasons to expect that the correct representation of our spacetime is given by the coordinate frame (\ref{spacetime basis vectors}). 

A constructive way of interpreting the constraint equation (\ref{constraint equation}) is to take it as a criterion to select the local frame. 

\medskip

We conjecture that the frame that correctly represents our $4D$ spacetime is that for which the condition $g_{\mu\nu,4}u^{\mu}u^{\nu} = 0$ is satisfied. Otherwise, the introduction of non-gravitational forces would be needed in order to keep the motion confined to spacetime \cite{Seahra3}, \cite{DynKK}. 

\medskip

Thus, from (\ref{usual equation of motion in 4D}) we conclude that observing the trajectories of  neutral test particles we would find no $5D$ effects to elucidate whether our spacetime is embedded in a world with more than four dimensions, like brane-world and induced matter theory.  This extends the classical results of Cho and Park \cite{Cho and Park}, to non-compact extra dimensions. 

\section{Effects From The Extra Dimension}

In this section we would like to comment on some observational/experimental implications of our work that can be used to distinguish the present theory from general relativity from an experimental point of view. Our discussion will be brief, with a view to inviting further in-depth study.

\subsection{Charge-To-Mass Ratio}

For a charged particle, its charge-to-mass ratio changes according to  (\ref{final expression for the charge to mass ratio}).  The first prediction is that $(q/m)$ varies even in the absence of electromagnetic field. In order to show this  effect, it is sufficient to consider the simplest case where the metric is independent of the extra coordinate. Equation (\ref{final expression for the charge to mass ratio}) can be integrated as
\begin{equation}
\label{charge-to-mass ratio}
\frac{1}{(q/m)^2} = - \frac{\epsilon}{\Phi^2} + C,
\end{equation}
where $C$ is a constant of integration. For astrophysical experiments/observations one should consider the metric for an isolated distribution of matter, which is pseudo Euclidean at spatial infinity. Assuming spherical symmetry
\begin{equation}
\label{astr. metric}
d{\bf {\cal S}}^2 = e^{\nu}(dt)^2 - e^{\lambda}[(dr)^2 + r^2(d\Omega)^2] + \epsilon \Phi^2 (d{\xi^4})^2,
\end{equation}
where $(d\Omega)^2 = (d\theta)^2 + sin^2\theta (d\phi)^2$, and the metric coefficients are some solution of the field equations. For example the Davidson and Owen solution \cite{Davidson and Owen}. The existence of the effects discussed here is independent of the specific details of the metric.

 The charge-to-mass ratio  (\ref{charge-to-mass ratio}) becomes
\begin{equation}
\label{q/m astrophysical case}
\left( \frac{q}{m}\right)^2 = \left(\frac{Q}{M}\right)^2 \frac{\Phi^2}{[\epsilon(Q/M)^2 + 1]\Phi^2 - \epsilon(Q/M)^2},
\end{equation}
where $(Q/M)$ is the mass-to-charge ratio measured at infinity ($\Phi(\infty) = 1$), i.e. by instruments not affected by gravity. We see that $(q/m)$ varies from its limiting value $(Q/M)$ at infinity to $(Q/M)[1 + \epsilon (Q/M)^2]^{-1/2}$ near the central object where $\Phi \gg 1$, which can be expected in the vicinity of black holes \cite{Davidson and Owen}. If the extra dimension is spacelike (timelike) then $(q/m)$ increases (decreases) as the particles moves towards the center.
\subsection{Motion of Charged Particles Vs. Motion of Neutral Particles }
The next prediction is, therefore, that the motion of a charged particle will differ from the motion of one without charge, even in the absence of electromagnetic field. Indeed, Eq.(\ref{equation of motion in terms of the ratio q/m}) indicates that a charged particle will be subjected to the force
\begin{equation}
\label{additional force}
f^{\mu} = \epsilon \left(\frac{q}{m}\right)^2 \frac{\Phi^{|(\mu)} - u^{\mu}\Phi_{|(\rho)}u^{\rho}}{\Phi^3},
\end{equation}
which vanishes for neutral particles. As a result of this, the locally measured radial velocity $V$ and the locally measured radial acceleration $g$ are different for neutral and charged particles. Namely, in coordinate frame,
\begin{eqnarray}
\label{V}
V^2 &=& 1 - \left(\frac{M}{E}\right)^2 e^{\nu}\left(\frac{Q/M}{q/m}\right)^2\nonumber \\
    &=& 1 - \left(\frac{M}{E}\right)^2 e^{\nu}\left[1 + \epsilon \left(\frac{Q}{M}\right)^2(1 - \Phi^{-2})\right],
\end{eqnarray}
where $E$ is the energy of the particle (at spatial infinity, $E = M/\sqrt{1 - V^2}$). The radial acceleration is 
\begin{equation}
\label{g}
g =\left[ -\frac{1}{2}\nu^{'} + \frac{(q/m)^{'}}{(q/m)}\right] e^{-\lambda/2}(1 - V^2),
\end{equation}
where prime denotes derivative with respect to $r$. For a neutral particle (\ref{V}) and (\ref{g}) reduce to the usual expressions in $4D$ general relativity \cite{JPdeL Weyl}. 

Thus, two different particles, one neutral and the other with electric charge, having the same mass and energy at infinity will be subjected to different accelerations and, therefore, develop different velocities. This effect could in principle be observed in experiments. 

\subsection{Inevitability of Peculiar Motion of Galaxies}
Let us now consider the motion of galaxies. In FRW universe models the locations of all galaxies are fixed by their comoving coordinates, which do not change as they recede from each other. But in the real universe they develop peculiar motions, in addition to the cosmic expansion. 

The study of peculiar motions could in principle allow us to detect the existence of extra dimensions. In fact, galaxies are neutral ``particles" and, therefore, the dependence of cosmological metrics on the extra coordinate (as in  \cite{JPdeL 1}) leads to the constraint equation (\ref{constraint equation}). For diagonal metrics it reduces to
\begin{equation}
\label{peculiar motion}
\frac{1}{g_{00}}\frac{\partial {g_{00}}}{\partial {\xi^4}} + 
(V^{i}_{pec})^2 \frac{\partial {g_{ii}}}{\partial {\xi^4}}= 0,
\end{equation}
which shows that, as a consequence of the dependence on the extra dimension, not all components of the spatial velocity can be zero simultaneously. In other words, galaxies cannot be fixed in space but necessarily have some peculiar motion with peculiar velocity $V_{pec}$. We stress the fact that this effect was missed in the ``old" Kaluza-Klein theory because of the imposition of the cylinder condition \cite{Cho and Park}.
\subsection{Variation of Thomson Cross Section and Fine Structure ``Constant"}

In a recent work \cite{QMinKK} we examined in more detail the effects of a large extra dimension on the rest mass and electric charge of test particles.  We showed that  both the rest mass and the charge vary along the trajectory observed in $4D$.  The constant of motion is now a combination  of these quantities. The possibility that these quantities  might be variable has important implications for the foundations of physics because variable mass and/or charge imply time-varying Thomson cross section  $\sigma = (8\pi/3)(q^2/m c^2)^2$ for the scattering of electromagnetic radiation by a particle of charge $q$ and mass $m$. This has been recognized by Hoyle \cite{Hoyle} and recently discussed from another viewpoint in \cite{Wesson and Seahra}.

Also the variation of electric charge $q$ implies the variation of the electromagnetic fine structure ``constant" $ \alpha_{em} = q^2/(4 \pi \hbar c)$. The latter has attracted considerable attention in view of the recent observational evidence that $\alpha_{em}$  might vary over cosmological time scales \cite{Webb1}-\cite{Webb2}. This, of course,  requires the time variation of at least one of the ``constants" ($q$, $\hbar$ and $c$). However, recently a number of theories attribute the variation of the fine structure constant to changes in the fundamental electron charge and preserve $c$ (Lorentz invariance) and $\hbar$ as constants \cite{Barrow1}-\cite{Youm4}.

Other consequences of the present formulation have been discussed in Refs.\cite{consequences 1} and \cite{consequences 2}. Finally, we note that the details (but not the existence) of the effects discussed here will depend on the specific model. This should give one the opportunity to test different models for their compatibility with observational and experimental data.

\section{Summary and Conclusions}
In this paper we have discussed the equation of motion of test particles for a version of Kaluza-Klein theory where the cylinder condition is not imposed. In this version, the metric tensor in $5D$ is allowed to depend on the fifth coordinate.

The equation of motion describing the trajectory of a particle as observed in $4D$ has been discussed in the context of space-time-matter theory and membrane theory. In these discussions the force (per unit mass) is defined as in equation (\ref{force in the literature}).  This force, which we call $f_{(lit)\mu}$, has a term parallel to the four-velocity of the particle. The existence of such force-term is a violation of four-dimensional laws of particle mechanics where $u_{\mu}f^{\mu} = 0$. Besides, this force can be finite or zero depending on the choice of coordinates and motion parameter. This is explicitly mentioned in references \cite{Youm1} and \cite{Seahra4}, and brings up the question of whether such abnormal force (or acceleration) is an effect from a large extra dimensions or it may be an spurious one due to wrong choice of frame or motion parameter.

This is disturbing because the results of physical observations generally do depend on the observer's frame of reference, but never on the particular set of coordinates, or parametrization, used. 

The advantage in this paper, with respect to other studies on the subject, is that at each step we make a clear difference between system of reference (defined by our choice of basis vectors) and system of coordinates.

Our main philosophy in this work may be summarized as follows.

\medskip

(i) The equations in $4D$ should be invariant under transformations in $5D$.
(ii) The metric tensor should lower and raise indexes, in such a way that covariant and contravariant components of a vector are simple related by $V_{\mu} = g_{\mu\nu}V^{\nu}$.
(iii) The (classical) theory should give a complete (deterministic) description of the motion of test particles.

\medskip

As a consequence of (i) we obtained that covariant derivatives in $4D$ should be calculated with the projected Christoffel symbols $\hat{\Gamma}^{\alpha}_{\mu\nu}$ defined in (\ref{projected Christoffel symbol}). In addition, we obtained the appropriate electromagnetic tensor invariant under $5D$ transformations. In the coordinate frame, it is given by (\ref{hatF}), which generalizes the usual one (\ref{Usual EM tensor}). 

In order to fulfill our condition (ii) we had to split the absolute derivative in such a way that the $4D$ covariant derivative of the metric tensor $g_{\rho\lambda}$, with respect to $\hat{\Gamma}^{\alpha}_{\mu\nu}$, vanishes. Then a new definition for the fifth force was proposed, such that $f_{\mu} = g_{\mu\nu}f^{\nu}$. This newly defined force turn out to be  always orthogonal to the four-velocity of the particle.  

As a consequence of our requirement (iii), that the equations completely specify the motion of the test particle, we identified the ``hidden" parameter (associated with the rate of motion along the extra dimension) with the electric charge, regardless of whether there is an electromagnetic field present or not. The appropriate general equations of motion were derived. 

It is important to note that the effects discussed in Section 7 are inevitable consequences of the assumed existence of extra dimensions. These effects should be observable, because they do not depend on the choice of coordinates or motion parameter. Their existence is model independent. However, the specific details will depend on the specific model. This should allow us to test different theoretical models with observational data.  

We would like to finish this paper with the remark that the general $4D$ equations of motion in an arbitrary spacetime frame $\hat{e}^{(\mu)}_{A}$ are given by (\ref{general expression for my definition of force, covariant}). Their particular version in the coordinate frame is given by (\ref{equation of motion in coordinate frame}). The validity of these equations is independent of the interpretation of $u^{(4)}$. Most probably, is better to work with them keeping $u^{(4)}_{0}$ as a free parameter. Thus leaving the possibility of different scenarios and interpretations. The theory discussed here can be easily extended to any number of dimensions.

\end{document}